\documentclass[prl,showpacs,twocolumn]{revtex4}

\usepackage{graphicx}

\begin{document}
\title{Feshbach Spectroscopy of a Shape Resonance}
\author{Thomas Volz$^1$, Stephan D{\"u}rr$^1$, Niels Syassen$^1$, Gerhard Rempe$^1$, Eric van Kempen$^2$, and Servaas Kokkelmans$^2$}
\affiliation{$^1$ Max-Planck-Institut f{\"u}r Quantenoptik, Hans-Kopfermann-Stra{\ss}e 1, 85748 Garching, Germany \\
$^2$ Eindhoven University of Technology, P.O.\ Box 513, 5600MB Eindhoven, The Netherlands}
%
\hyphenation{Fesh-bach}
\begin{abstract}
We present a new spectroscopy technique for studying cold-collision properties. The technique is based on the association and dissociation of ultracold molecules using a magnetically tunable Feshbach resonance. The energy and lifetime of a shape resonance are determined from a measurement of the dissociation rate. Additional spectroscopic information is obtained from the observation of a spatial interference pattern between an outgoing $s$ wave and $d$ wave. The experimental data agree well with the results from a new model, in which the dissociation process is connected to a scattering gedanken experiment, which is analyzed using a coupled-channels calculation.
\end{abstract}
\maketitle

The field of ultracold molecules has seen tremendous progress in the past two years. The key technique that triggered this development is the association of ultracold molecules from ultracold atoms using a Feshbach resonance \cite{regal:03,herbig:03,duerr:04,strecker:03,cubizolles:03,jochim:03a,mukaiyama:04,zwierlein:04,thompson:05}. A Feshbach resonance is caused by the resonant coupling of a colliding atom pair to a molecular bound state. The system can be tuned into resonance by applying a magnetic field. To produce stable molecules, the magnetic field is slowly ramped across the Feshbach resonance to lower fields. For detection, the molecules are dissociated by ramping the magnetic field back to higher values above the Feshbach resonance. In this field range, the energy of the molecular state lies in the continuum of unbound atom pair states. The coupling between the discrete molecular state and the continuum of atomic states leads to an exponential decay of the molecules into unbound atom pairs. 
These atoms are then imaged with standard techniques. Beyond this molecule detection scheme, the dissociation has other interesting applications: For example, the width of the Feshbach resonance can be extracted from the energy released in the dissociation process \cite{mukaiyama:04,duerr:04a}.

In this letter, we use the dissociation process as a new spectroscopic tool for studying cold-collision properties. Specifically, we investigate a $d$-wave shape resonance in $^{87}$Rb. The shape-resonance state is a quasi-bound state which is localized behind the centrifugal barrier. The energy of this state lies in the continuum of unbound atom pair states. In our experiment, the shape resonance is probed by tuning the energy of the Feshbach molecules with the magnetic field. If the energy of the molecules matches the energy of the shape resonance, then population is transferred from the molecular state to the shape-resonance state. From there, the population decays rapidly into unbound atom pairs by tunneling through the centrifugal barrier. The molecules can therefore dissociate in two ways, either directly into the continuum or indirectly by passing through the shape-resonance state. The indirect processes turn out to be much faster than the direct ones, so that the molecule dissociation rate is drastically enhanced. However, if the energies of the molecules and the shape resonance do not match, then indirect processes will be suppressed. The magnetic-field dependence of the dissociation rate thus reveals the energy and lifetime of the shape resonance. This information can be used as spectroscopic input to constrain theoretical models for the cold-collision properties. Little is gained in the case of $^{87}$Rb, where the collisional properties are rather well known, and where the shape resonance has been observed in previous experiments \cite{boesten:97,thomas:04,buggle:04}. However, our method is interesting for systems, where the collisional properties are unknown, such as heteronuclear mixtures or non-alkali species. Here, theoretical models starting from ab-initio calculations presently do not have enough experimental input to obtain realistic predictions for the cold-collision properties.

The experimental setup was described in detail elsewhere \cite{marte:02,duerr:04,duerr:04a}. In brief, a Bose-Einstein condensate (BEC) of $^{87}$Rb atoms in the hyperfine state $|f,m_f\rangle = |1,1\rangle$ is prepared in an optical dipole trap. The magnetic field is held slightly above the Feshbach resonance, which is located at $B_{res} \sim 632$~G \cite{marte:02}. It is caused by a $d$-wave molecular state and has a width of 1.3~mG \cite{duerr:04a}. The BEC is released from the trap and 1~ms later molecules are created by slowly ramping the magnetic field downward across the Feshbach resonance. Remaining atoms are spatially separated from the molecules using a Stern-Gerlach field. Next, the molecules are dissociated back into unbound atom pairs by jumping the magnetic field to a value $B>B_{res}$ and holding it there for a variable time $t_{hold}$ \cite{note:technical-detail}. During $t_{hold}$, population in the molecular state decays exponentially with a rate $\Gamma(B)$, which depends on the value of $B$ during $t_{hold}$. After $t_{hold}$, the  magnetic field is switched off rapidly, which stops the dissociation process. After 0.85~ms time of flight (counting from the dissociation), an absorption image is taken. Molecules that did not decay during $t_{hold}$ are invisible in the image, because only unbound atoms resonantly absorb light from the detection laser beam.

\begin{figure} [bt]
\includegraphics[width=.35\textwidth]{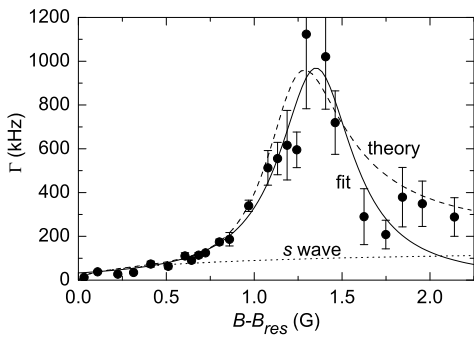}
\caption{
 \label{fig-total-rate}
Molecule dissociation rate $\Gamma$ as a function of magnetic field $B$ with respect to the position of the Feshbach resonance $B_{res}$. The experimental data for the total rate (circles) clearly show the effect of the shape resonance near 1.3~G. The solid line is a Lorentzian fit to the data. The dashed (dotted) line shows the prediction for the total ($s$-wave) dissociation rate from a coupled-channels calculation. The $s$-wave dissociation is not affected by the shape resonance.
 }
\end{figure}

The dissociation rate $\Gamma(B)$ is extracted from a series of images for fixed $B$ and variable $t_{hold}$. The number of dissociated atoms is determined from each image and then fit to an exponential as a function of $t_{hold}$. This yields the dissociation rate shown in Fig.~\ref{fig-total-rate}. The enhanced dissociation rate near 1.3~G due to the shape resonance is clearly visible.

\begin{figure} [bt]
\includegraphics[width=.4\textwidth]{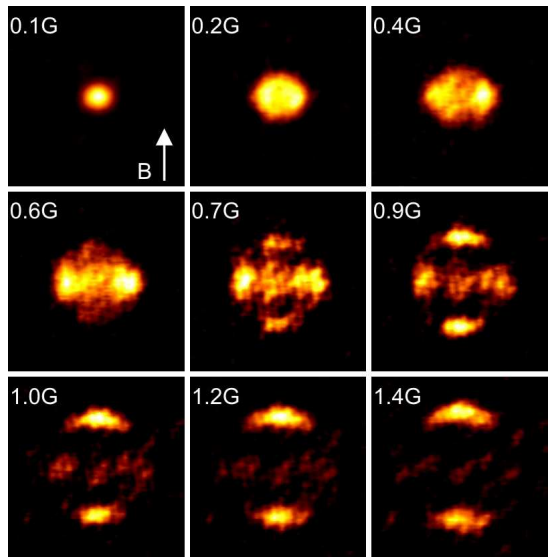}
\caption{
 \label{fig-2D-images} 
(Color online) Time-of-flight images of unbound atoms obtained in the molecule dissociation. The images were taken at different values of $B-B_{res}$. The magnetic field is vertical in the images. The interference between the $s$ and $d$ partial wave undergoes a change in relative phase and amplitude. At 0.1~G, the dissociation is mostly $s$ wave, producing a circle. For higher magnetic fields, both partial waves are populated. From 0.2 to 0.6~G, the interference suppresses atom emission {\em along} $B$, whereas the opposite relative phase in the interference between 1.0 and 1.4~G suppresses emission {\em perpendicular} to $B$. At 0.7 and 0.9~G, the relative phase is such that neither component is strongly suppressed. The typical radius reached by the atoms during the constant time of flight increases for increasing $B$, thus indicating an increase of the kinetic energy released in the dissociation.
 }
\end{figure}

In a simple analytic model, the dissociation can be described as a two-step process. In the first step, the interaction Hamiltonian $H$ transfers population from the molecular state $|\psi_{mol}\rangle$ to the shape-resonance state $|\psi_{shape}\rangle$. This is described by a generalized Rabi frequency $\Omega = (2/\hbar) |\langle \psi_{shape}|H| \psi_{mol}\rangle|$. In the second step, the population tunnels with a rate $\Gamma_{shape}$ from $|\psi_{shape}\rangle$ into the continuum of unbound atom pair states. One can show that for $\Omega \ll \Gamma_{shape}$ the molecular state decays exponentially with a rate \cite{cohen-tannoudji:92:p49}
\begin{eqnarray}
 \label{eq-Gamma-B}
\Gamma(B) = \frac{\Omega^2}{\Gamma_{shape}} \; \left[1+ \left( 2 \; \frac{E_{shape} - E_{mol}(B)} 
{\hbar\Gamma_{shape} }\right)^2 \right]^{-1} ,
\end{eqnarray}
where $E_{shape}$ and $E_{mol}$ are the energies of the shape-resonance state and the molecular state, respectively. $E_{mol}$ depends on $B$, while $\Gamma_{shape}$, $E_{shape}$, and $\Omega$ in this simple model do not. For the small magnetic-field range considered here, a linear approximation holds
\begin{eqnarray}
 \label{eq-E-F}
E_{mol}(B)=(B-B_{res})\Delta\mu \; ,
\end{eqnarray}
where $B_{res}$ is the position of the Feshbach resonance and $\Delta \mu$ is the difference in the magnetic moments between a molecule and an unbound atom pair. $\Delta \mu$ is determined experimentally from the time-of-flight images shown in Fig.~\ref{fig-2D-images}. The radial cloud size in the images yields the kinetic energy $E$ in the relative motion of the dissociated atom pairs. Due to energy conservation in the dissociation process, $E = E_{mol}(B)$. We thus obtain $\Delta \mu = k_B \times 230(18)~\mu$K/G. Alternatively, $\Delta \mu$ can be determined with a Stern-Gerlach method \cite{duerr:04}.

$\Gamma(B)$ in Eq.~(\ref{eq-Gamma-B}) is a Lorentzian, which is fit to the data as shown in Fig.~\ref{fig-total-rate}. Using the above value for $\Delta \mu$, the best-fit parameters are $E_{shape}=k_B \times 312(25)~\mu$K, $\Gamma_{shape}=15(3)$~MHz, and $\Omega=2\pi\times 0.61(7)$~MHz. We thus extract the energy and lifetime of the shape resonance as well as the Rabi frequency without making any reference to the coupled-channels calculation 
further below.

Figure \ref{fig-2D-images} shows time-of-flight images of the atoms after dissociation. They exhibit spatial interference patterns created by different outgoing partial waves. Only two partial waves contribute to these interference patterns, namely those with quantum numbers $(l,m_l) = (0,0)$ and $(2,0)$ for the rotation of the atoms around each other \cite{note:selection-rules}. The indirect dissociation processes (where population passes through the shape-resonance state) create outgoing $d$ waves, because the shape-resonance state is a $d$-wave state and the tunneling out of this state does not affect the angular momentum. In contrast, the direct dissociation processes are independent of the shape-resonance and preferentially populate the $s$-wave state, because all other partial waves are suppressed for low energy by the centrifugal barrier. Note that interference between different partial waves of cold atoms has previously been observed in {\em scattering} experiments \cite{legere:98,thomas:04,buggle:04}. Previous molecule {\em dissociation} experiments, however, observed only outgoing $s$ waves.

The interference pattern is described by the following wave function
\begin{eqnarray}
 \label{eq-psi-decay}
\psi_{decay} (\vec r) = g(r,t) \left( \sqrt{\beta_0} Y_{00} - e^{i\delta_{rel}} \sqrt{\beta_2} Y_{20}(\vartheta) \right)
\end{eqnarray}
where $Y_{lm}(\vartheta,\varphi)$ is a spherical harmonic, the real numbers $\beta_l$ and $\delta_{rel}$ characterize the amplitudes of the partial waves and their relative phase, and $g(r,t)$ is a radial wave function. $g(r,t)$ is normalized such that $\beta_0 +\beta_2=1$, which implies that $\beta_l$ is the branching ratio for decay into the $l$-th partial wave.

\begin{figure} [bt]
\includegraphics[width=.35\textwidth]{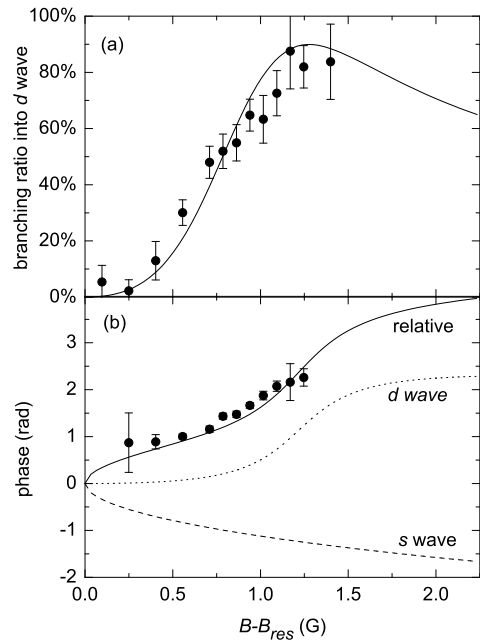}
\caption{
 \label{fig-branching-ratio-and-phase} 
Parameters extracted from the spatial interference patterns. (a) Branching ratio for decay into the $d$ wave. Again, the $d$-wave shape resonance is clearly visible. (b) Partial-wave phases and relative phase. The solid, dashed, and dotted lines show the theoretical prediction for the relative phase $\delta_{rel}= \delta_2^{bg} -\delta_0^{bg}$, the $s$-wave phase $\delta_0^{bg}$, and the $d$-wave phase $\delta_2^{bg}$, respectively. Experimental data (circles) agree well with the theory (solid line) in (a) and (b).
 }
\end{figure}

The branching ratio $\beta_2$ and the relative phase $\delta_{rel}$ are extracted from the images with the same method as used in Ref.~\cite{buggle:04}. In brief, computed tomography is used to reconstruct the 3-dimensional (3D) density distribution from the 2D images. The 3D density is sorted into 20 bins to obtain the probability $W(\vartheta)$ for finding an atom at angle $\vartheta$. A fit of the angular part of the modulus squared of Eq.~(\ref{eq-psi-decay}) to $W(\vartheta)$ yields the fit parameters $\beta_2$ and $\delta_{rel}$. These parameters are shown in Fig.~\ref{fig-branching-ratio-and-phase}. The branching ratio clearly shows the enhanced decay into the $d$ wave due to the shape resonance. For $\beta_2$ close to 0 or 1, the fit cannot reliably determine the relative phase.

In order to analyze the dissociation more rigorously, we performed a coupled-channels calculation for a scattering {\em gedanken} experiment that is closely related to our dissociation experiment. We briefly summarize the theory here. For more details see Ref.~\cite{duerr:prep}. In the {\em gedanken} experiment, we consider a colliding atom pair with kinetic energy $E$ in the relative motion in the presence of a magnetic field $B$ near the 632~G Feshbach resonance. $S$-matrix elements $S_{ll'}$ are calculated in the $E$-$B$-plane for $l$ and $l'$ equal to 0 or 2 (both with $m_l=0$). According to the theory of multichannel scattering resonances, these $S$-matrix elements are expected to follow an analytic expression of the form
\cite{taylor:72,feshbach:92}
\begin{eqnarray}
 \label{eq-S-matrix}
S_{ll'}(E,B) & = & \left( \delta_{ll'} - \frac{i\hbar\Gamma_{ll'}(E)}{E-E_{mol}(B)+\frac{i}{2}
\hbar\Gamma(E)} \right) \nonumber \\
 && \times e^{i[\delta_l^{bg}(E) + \delta_{l'}^{bg}(E)]} \; .
\end{eqnarray}
Here, $\delta_{ll'}$ is the Kronecker symbol, $\Gamma(E)$ is the total decay rate of the molecular state, $E_{mol}(B)$ is given by Eq.~(\ref{eq-E-F}), and the significance of the partial-decay-rate parameters $\Gamma_{ll'}(E)$ will be discussed below. Finally, $\delta_l^{bg}(E)$ is the background scattering phase for the $l$-th partial wave, where background means for magnetic fields far away from the Feshbach resonance \cite{note:S-matrix}.

For any given $E$, we fit Eq.~(\ref{eq-S-matrix}) for variable $B$ to the $S$-matrix elements obtained from our coupled-channels calculation. The fits match the $S$-matrix extremely well. The open-channel physics, including the shape resonance, is  independent of $B$ and is therefore included in the energy-dependent fit parameters $\delta_l^{bg}(E)$ and $\Gamma_{ll'}(E)$. Furthermore, the combination of the fits yields $\Delta \mu = k_B \times 224~\mu$K/G.

We will now discuss the connection between our dissociation experiment and the fit parameters $\delta_l^{bg}(E)$ and $\Gamma_{ll'}(E)$. As mentioned earlier, the kinetic energy $E$ of the atom pairs after dissociation is $E=E_{mol}(B)$. Therefore the energy-dependent fit-parameters for the {\em gedanken} experiment become magnetic-field dependent parameters in the dissociation experiment $\delta_l^{bg}(B)$ and $\Gamma_{ll'}(B)$.

The link between the scattering {\em gedanken} experiment and the dissociation experiment can be established from evaluating the different terms in the asymptotic form of the regular scattering wave function
\begin{eqnarray}
\psi^{(+)}(\vec r) & \stackrel{r \rightarrow \infty}{\longrightarrow} & (-1)^{l'} \frac{e^{-ikr}}{r} \; Y_{l'0} (\vartheta)  \nonumber \\ &&
 - \; \frac{e^{ikr}}{r} \sum_{l} \left( S^{bg}_{ll'} + S^{res}_{ll'} \right) Y_{l0} (\vartheta) \; ,
\end{eqnarray}
where we assume in the {\em gedanken} experiment that only one partial wave $l'$ is initially populated. Here, $k$ is related to the energy by $E=\hbar^2k^2/(2m_{red})$ with the reduced mass $m_{red}$, and we split the $S$-matrix into a background part and a resonant part $S_{ll'}= S^{bg}_{ll'} + S^{res}_{ll'}$ with $S^{bg}_{ll'}=e^{2i\delta_l^{bg}(E)} \delta_{ll'}$.

The scattering wave function consists of three parts: an incoming wave, an outgoing background wave, and an outgoing resonant wave. The resonant part is due to particles that make the transition to the molecular state and subsequently decay back into the open channel. The Stern-Gerlach separation removes all incoming flux and along with it the background scattered wave. Hence, these two terms need to be removed from the scattering state in order to
describe the decay wave function
\begin{eqnarray}
\psi_{decay}(\vec r) & \stackrel{r \rightarrow
\infty}{\longrightarrow} &
 - \; \frac{e^{ikr}}{r} \sum_{l} S^{res}_{ll'} Y_{l0} (\vartheta) \; .
\end{eqnarray}
Using the above equations, this can be rewritten to yield Eq.~(\ref{eq-psi-decay}), where the relative phase of the decaying partial waves is $\delta_{rel}= \delta_2^{bg}- \delta_0^{bg}$ and the branching ratio for decay into the $l$-th partial wave is given by $\beta_l(B) = \Gamma_{ll}(B)/ \Gamma(B)$. This means that $\Gamma_{ll}(B)$ represents the partial decay rate into the $l$-th partial wave. The theory for the decay rate, branching ratio, and the phases are shown as lines in Figs.~\ref{fig-total-rate} and \ref{fig-branching-ratio-and-phase}. The good agreement between theory and experiment without any free fit parameters supports the theory developed here.

Finally, we extract the parameters of the simple model from the full theory. The $d$-wave decay rate $\Gamma_{22} = \Gamma - \Gamma_{00}$ is the difference between the total and the $s$-wave decay rate, both shown in Fig.~\ref{fig-total-rate}. The maximum of $\Gamma_{22}$ is located at 1.28~G corresponding to $E_{shape} = k_B \times 287~\mu$K. The value and the curvature of $\Gamma_{22}(B)$ at the maximum correspond to $\Gamma_{shape} = 17$~MHz and $\Omega=2\pi\times 0.61$~MHz. These numbers agree well with the values obtained by fitting the simple model Eq.~(\ref{eq-Gamma-B}) to the experimental data.

In conclusion, we showed that dissociation of ultracold molecules can be used to measure the energy and lifetime of a shape resonance. Additional information was obtained from the analysis of the spatial interference patterns. This is enough information to constrain theoretical models so much that the basic scattering properties of the system can be determined. We developed a new model for the dissociation, which agrees well with the experimental results.

We acknowledge fruitful discussions with Harald Friedrich and Boudewijn Verhaar. This work was supported by the European-Union Network ``Cold Quantum Gases". S.K.\ acknowledges support from the Netherlands Organization for Scientific Research (NWO). E.K.\ acknowledges support from the Stichting FOM, financially supported by NWO.


\end{document}